\documentclass[]{spie}  %>>> use for US letter paper
\usepackage{amsmath,amsfonts,amssymb}
\usepackage{graphicx}
\usepackage{comment}
\usepackage{hyperref}
\hypersetup{colorlinks=true, allcolors=blue}
\usepackage{tabularx}

\title{Realizing a robust, reconfigurable active quenching design for multiple architectures of single-photon avalanche detectors}

\author[a]{Subash Sachidananda}
\author[c]{Prithvi Gundlapalli}
\author[c]{Victor Leong}
\author[c]{Leonid Krivitsky}
\author[a,b]{Alexander Ling}
\affil[a]{Centre for Quantum Technologies, National University of Singapore, 3 Science Drive 2, Singapore 117543
}
\affil[b]{Department of Physics, National University of Singapore, 2 Science Drive 3, Singapore 117551}
\affil[c]{Institute of Materials Research and Engineering, Agency for Science, Technology and Research (A*STAR), Singapore
}

\authorinfo{For further information on active quench system: Subash Sachidananda - E-mail: subash.jies@gmail.com\\ For further information on custom fabricated APD: Victor Leong - E-mail: victor\_leong@imre.a-star.edu.sg}

\begin{document} 
\maketitle

\begin{abstract}
Most active quench circuits used for single-photon avalanche detectors are designed either with discrete components which lack the flexibility of dynamically changing the control parameters, or with custom ASICs which require a long development time and high cost. As an alternative, we present a reconfigurable and robust hybrid design implemented using a System-on-Chip (SoC), which integrates both an FPGA and a microcontroller. We take advantage of the FPGA’s speed and configuration capabilities to vary the quench and reset parameters dynamically over a large range, thus allowing our circuit to operate with a wide variety of APDs without having to re-design the system. The microcontroller enables the remote adjustment of control parameters and re-calibration of APDs in the field. The ruggedized design uses components with space heritage, thus making it suitable for space-based applications in the fields of telecommunications and quantum key distribution (QKD). We characterize our circuit with a commercial APD cooled to -20°C, and obtain a deadtime of 35ns while maintaining the after-pulsing probability at close to 3$\%$. We also demonstrate versatility of the circuit by directly testing custom fabricated chip-scale APDs, which paves the way for automated wafer-scale testing and characterization.
\end{abstract}

\keywords{single photon avalanche detectors , System on chip SoC, FPGA, active quenching, chip-scale APD, reconfigurable quenching circuit, automated wafer-scale testing}

\section{INTRODUCTION}
Single photon detection is used widely in quantum optics technology. Increasing the rate and efficiency of single photon detection is a fundamental way to improve the overall efficiency of optical systems. The most popular method to detect single photons is by using Geiger-mode Avalanche Photo Diodes (GM-APD) in which photon absorption triggers a current due to avalanche multiplication process. This avalanche current needs to be stopped quickly and the APD needs to be restored to its normal operating mode before it can detect the next photon. This process is called $quenching$ and can be done either passively or actively. Although $passive\ quench$ (PQ) \cite{Cova96, wincomparator} circuits are simple to implement they do not reset the diode quickly to its nominal state which limits their detection efficiency. $Passive\ Quench\ Active\ Reset$ (PQAR) \cite{pqar} circuits can provide shorter and well-defined reset times but fail to restrict the after-pulsing probability of the APD within acceptable bounds if the reset times set are too short. $Active\ quench$ (AQ) \cite{Gallivanoni10} circuits on the other hand provide fast response and reset times while limiting the after-pulsing probability of the APDs and hence are most suitable for high detection rate applications.

\subsubsection*{Motivation}
In several active quenching circuit designs, generation of the quench and reset signals for the APD is done using only discrete hardware components \cite{Ghioni96, Stipcevic09, Stipcevic17}. Such designs are cheaper to manufacture but critical parameters such as the quench and reset times are hard bound and cannot be changed dynamically. Changing the quench and reset time can compensate for deviation in APD performance at run-time occurring due to factors such as temperature variation. As an alternative to discrete systems, monolithic integrated \cite{zappa2000, Acconcia16, 37ps-aq, Ceccarelli2019, Zimmerman5ns} active quench circuits have been developed which offer flexibility to change parameters dynamically while also reaching very high count rates and detection efficiency. However fabricating custom ICs is a complex process requiring long development cycles and a very high manufacturing cost. 
%It also creates geographical dependencies on specific fabrication houses which makes the process commercially less viable.

In this paper we describe a hybrid design for active quenching of Geiger-Mode APDs (GM-APD) comprising of minimal discrete components and a commercially available System-on-Chip (SoC) \cite{zedboard} which has both an FPGA and a micro-controller integrated on the same chip. The active quench control logic is implemented on the FPGA which allows to dynamically change typical active quench parameters such as quench width, reset width, deadtime, etc. The microcontroller interfaces with a PC and provides remote access to the FPGA registers which allows to re-calibrate the APDs deployed in the field during run-time. Only a few discrete components need to be placed close to the APD head while the SoC can be placed farther away without compromising the performance and detection efficiency. We present results to show that our system can achieve a deadtime of 35ns despite the APD head located $\approx$15cm away from the SoC, thus allowing to reach count rates of $>$28Mcps. Although our system is mainly designed for active quenching, we show how it can also be reconfigured to work in either passive quench or PQAR configurations.

%At a system level this design has several advantages. It reduces the overall real estate area of the PCB close to the APD head making it potentially suitable for size/volume-constrained applications with strict SWAP (size-weight-area-power) budgets, such as nano-satellites. The use of the microcontroller allows for unique identification and access to each APD attached to the circuit through a connected network such as RF wireless and/or Ethernet. 
%It also enables to run diagnostic self-calibrating algorithms and dynamically correct for drifts in APD performance. This is useful for APDs which are deployed in harsh environments where human access is either difficult or impossible. 

%Our design supports a wide range of values for the tune-able parameters which provides flexibility to directly use it with many APDs from several manufacturers without modifications. 

The tunable parameters in our design can be varied over a wide range allowing operation with APDs from several manufacturers without modifications. We demonstrate this versatility by simultaneously operating our system with a commercial APD \cite{sap500} and our in-house custom fabricated chip-scale APD \cite{imreAPD21}, both of which have different characteristics in terms of breakdown voltage, over-voltage, etc. We present some preliminary results for breakdown voltage and dark counts obtained with the custom fabricated chip-scale APDs which further validates our design. In principle, the flexibility in our system would be useful both in a research setup comprising of multiple device variants and in mass production for doing direct wafer-scale testing. 

%This is especially true for future testing with our custom chip-scale APDs, for which there exist several hundred variants with each one different from the other in terms of channel length, width, doping concentration, etc. Characterizing each of these variants for multiple performance metrics such as dark count, after-pulsing, jitter, linearity etc. would involve dynamic adjustments of several active quench circuit parameters (quench/reset duration, deadtime, bias/quench voltages, etc.). This results in an exponential increase in the total number of test cases and doing them using hard-wired fixed active quench circuits is impractical since it involves a lot of manual intervention which is both laborious and error prone. Using our system instead paves way for automation of such testing and characterization thus saving cost and time.
\section{Design and implementation}
\label{sec:design}
\begin{figure} [ht]
\begin{center}
\begin{tabular}{c} 
\includegraphics[width=\textwidth]{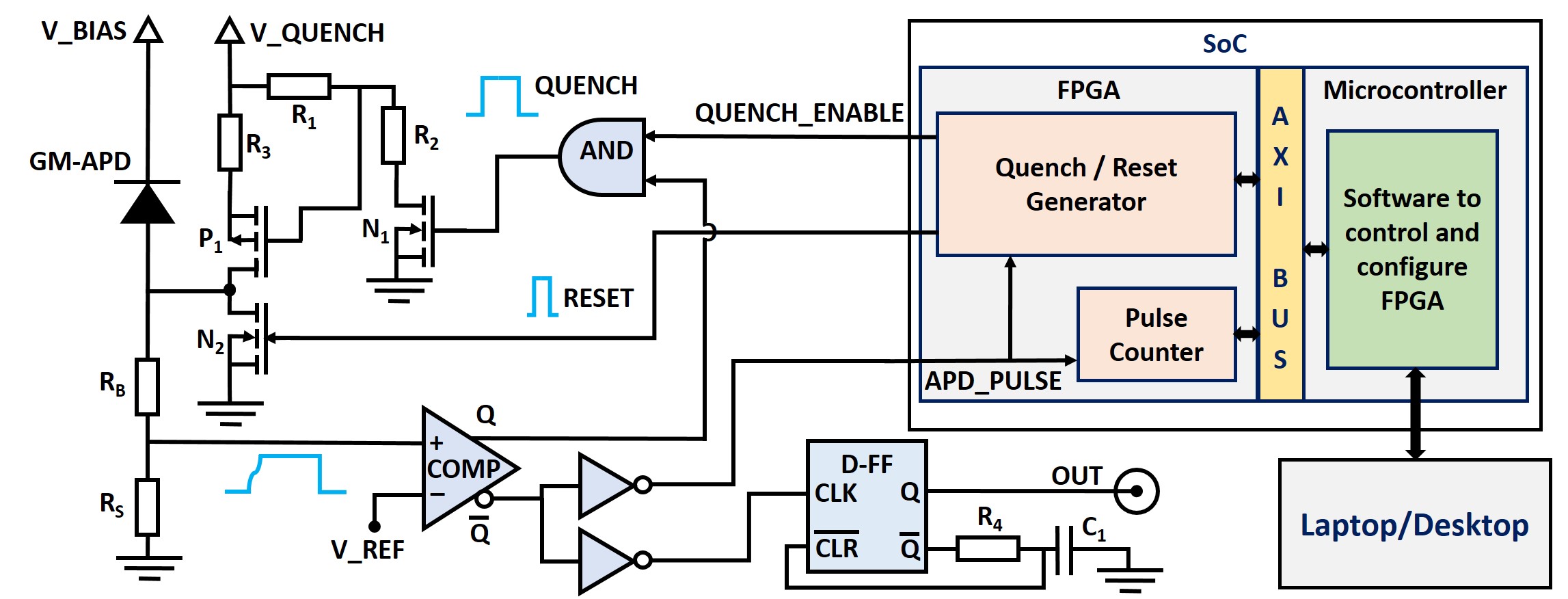}
\end{tabular}
\end{center}
\caption[] 
{ \label{aq-design} 
Simplified circuit diagram of the system with Geiger Mode APD (GM-APD)}
\end{figure}

Similar to conventional active quenching circuits, our system operates in two phases in response to the occurrence of an avalanche: 1) $quench$ - reduce the bias voltage of the diode below its breakdown voltage for a few nanoseconds and 2) $reset$ - restore the bias voltage above the breakdown for a few nanoseconds while bypassing the ballast resistor at the same time. After the reset phase, the ballast resistor is reconnected and the APD returns to the nominal state. The main functional blocks of our system are shown in figure \ref{aq-design}. The $ZedBoard$ \cite{zedboard} development kit with a Zynq-7000 based SoC is used for realizing the FPGA and microcontroller modules. The GM-APD is connected with the quench/reset generator on the FPGA using a custom PCB with minimal discrete components. We use the LT6752-3 high speed comparator (COMP) which produces complementary outputs after detecting the avalanche pulse across sense resistor $R_S$. The feedback path for doing the quench comprises of a fast AND gate, P-MOSFET $P_1$ and N-MOSFET $N_1$. Reset is done through N-MOSFET $N_2$. The values of $R_4$ and $C_1$ are set such that the D-FlipFlop (D-FF) trims the comparator output to 20ns pulses (OUT) which are used for characterizing the deadtime and after-pulsing probability. Matsusada high voltage DC-DC converters are used to derive a bias voltage (V\_BIAS) between 0-500V while quench voltage V\_QUENCH is derived from an external power supply.
\begin{figure} [ht]
\begin{center}
\begin{tabular}{c} 
\includegraphics[width=0.65\textwidth]{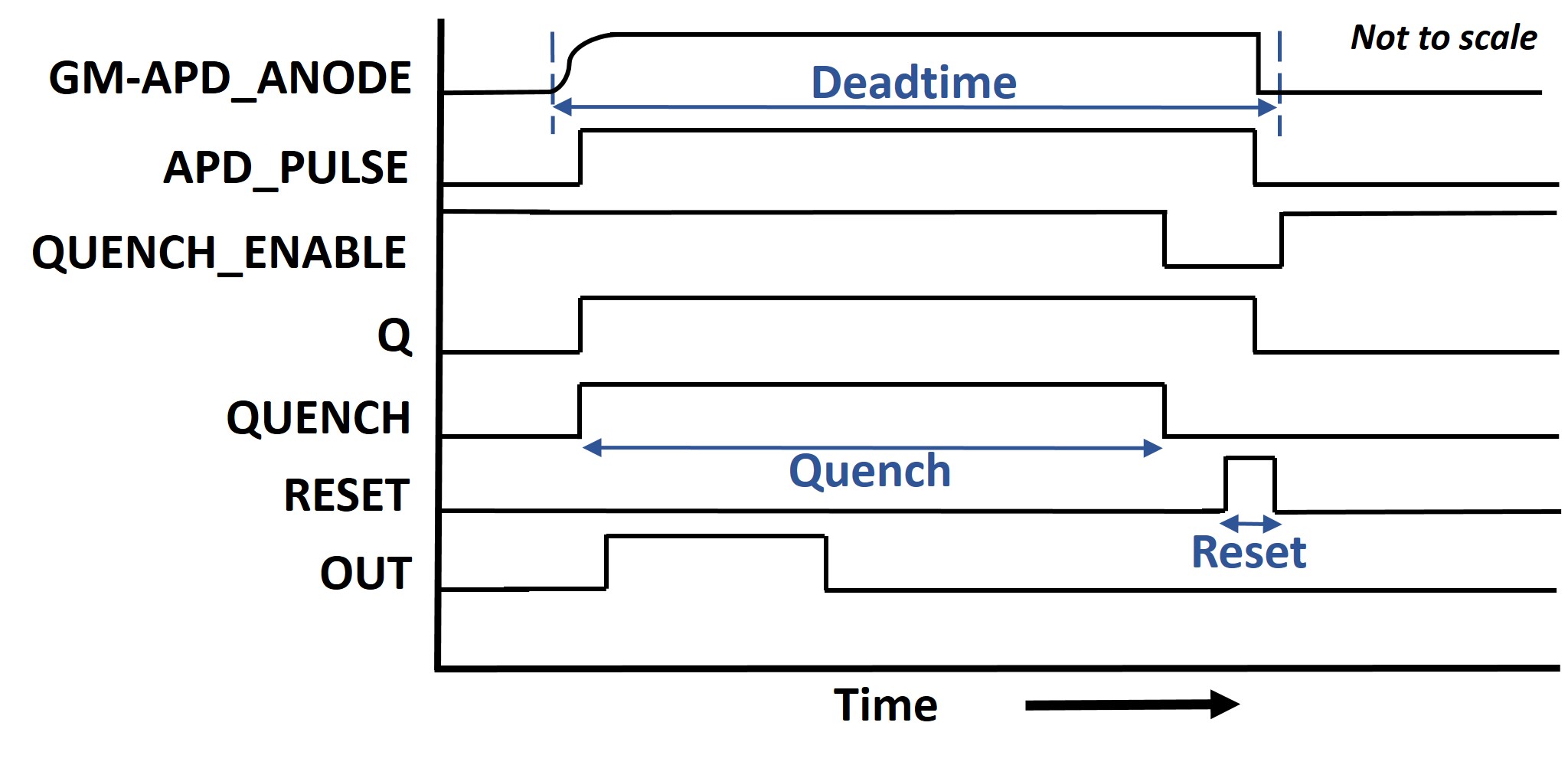}
\end{tabular}
\end{center}
\caption[] 
{ \label{aq-timingdiagram} 
Timing diagram of critical signals showing the active quench operation after occurrence of an avalanche}
\end{figure}
\subsection{Operation in active quench configuration}
\begin{figure} [ht]
\begin{center}
\begin{tabular}{c} 
\includegraphics[width=\textwidth]{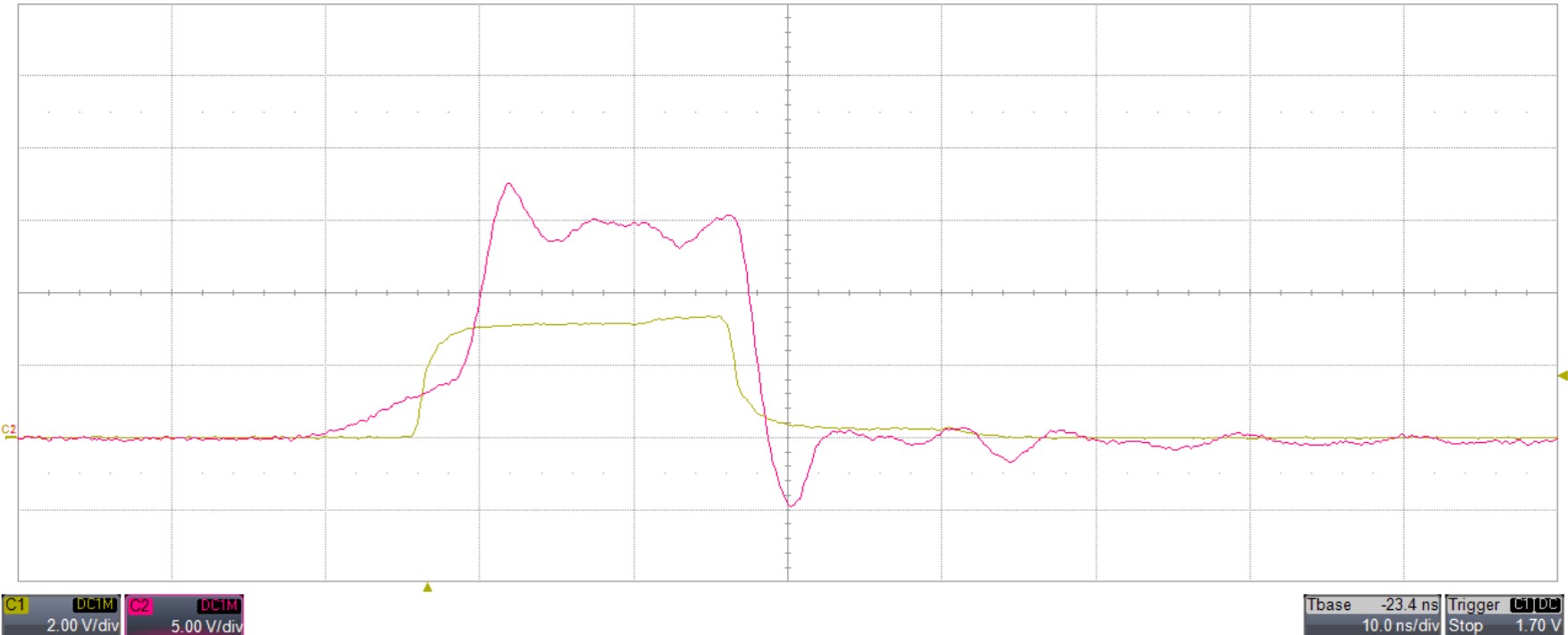}
\end{tabular}
\end{center}
\caption[] 
{ \label{fig:aq-anode} 
Waveform of the OUT pulse (C1) and GM-APD\_ANODE voltage (C2) captured on the oscilloscope}
\end{figure}
The design and operation is better understood with the timing diagram shown in figure \ref{aq-timingdiagram}. The avalanche current in the APD causes the voltage at the APD anode (GM-APD\_ANODE) to rise and consequently the voltage across $R_S$ exceeds the comparator reference voltage (V\_REF), which is typically set to $\approx$17mV. In response the comparator (COMP) toggles the complementary output pulses $Q$ to high and $\overline{Q}$ to low. Output $Q$ is fed back to one of the inputs of the AND gate while the FPGA keeps the other input of the AND gate (QUENCH\_ENABLE) high before the avalanche occurs. So the AND gate output (QUENCH) goes high to turn ON $N_1$ which subsequently turns ON $P_1$ to initiate quenching. As a result of this, the anode voltage of GM-APD reaches the quench voltage (V\_QUENCH) which effectively reduces the voltage across the APD below its breakdown voltage and brings it into the quench state. Until then the APD is passively quenched and the avalanche current is limited by the ballast resistor ($R_B$). This feedback path is kept as short as possible to reduce the response time and after-pulsing probability\cite{Wayne-AP}.
%Rb=43k and Rs=1k

The other comparator output $\overline{Q}$ is inverted (APD\_PULSE) and fed into the quench/reset logic to register the occurrence of an avalanche and into a 24-bit counter to record the total number of avalanches per second. The duration for which the APD stays in the quench state is configurable and after this duration the FPGA sets the QUENCH\_ENABLE signal to low, as shown in figure \ref{aq-timingdiagram}. This drives the output of the AND gate to low which turns OFF $N_1$ and $P_1$. The FPGA then initiates the reset state by setting the RESET signal high to turn ON $N_2$. The GM-APD\_ANODE voltage then dips to 0 which brings the relative bias voltage of the APD above its breakdown voltage. This also makes the comparator output $Q$ to go low and $\overline{Q}$ to go high after which $N_2$ is turned OFF by setting the RESET signal low. The duration of reset is configurable and typically kept as short as possible. To ensure that both $P_1$ and $N_2$ are not ON at the same time, the FPGA inserts a tiny configurable delay between end of the quench and start of the reset. The total duration from the start of the avalanche to the completion of the reset constitutes the $deadtime$ of the system. After reset the QUENCH\_ENABLE signal is driven high again to keep the system ready for the next detection event.

%The microcontroller software can directly configure registers of the FPGA modules as memory mapped entities over the standard AXI bus. For every parameter/register/flag in each FPGA module, there is a corresponding variable in the software. These can be initialized and changed by the microcontroller software at run-time. The microcontroller also provides a bi-directional $command-response$ interface and connects with a PC (laptop/desktop) using a serial communication port. So the active quench parameters (such as quench, reset, delay, deadtime, counts, etc.) can be accessed during the experiment from the PC either directly through user-entered commands or by automated scripts which can read and parse these commands from a configuration file. This also enables for high precision monitoring and control of all APDs used during the experiment to ensure that the parameters like quantum bit error rate (QBER) is within agreeable limits at all times.

%In addition, the microcontroller software periodically reads the pulse counter on the FPGA to keep track of the number of detection events per second. This is further used for characterization parameters such as dark counts, linearity, etc. of the APD. A copy of the APD\_PULSE is also fed into a separate (off-chip) pulse trimming circuit comprising of a D-Flip Flop  These pulses can be either fed into an oscilloscope  

The operation of our system is verified with a commercial SAP500-TO8 APD \cite{sap500}, biased with an excess voltage of 10V. V\_QUENCH is set to 15V, which potentially allows to operate the APD up to an excess voltage 12V. Figure \ref{fig:aq-anode} shows the GM-APD\_ANODE voltage (C2) captured with a LeCroy Waverunner 610Zi oscilloscope. After the avalanche starts, the activation of quench occurs at $\approx$9ns. Referring to figure \ref{aq-design}, this delay is incurred in the quench feedback path which is the sum of propagation delays through the comparator, AND gate and the switch ON delays for $N_1$ and $P_1$. After quench is activated, the GM-APD\_ANODE voltage rises rapidly to V\_QUENCH within 2ns. The OUT signal (C1) is delayed by $\approx$6ns due to the propagation delays through the comparator, inverting buffer and D-Flipflop (D-FF). 

The minimum quench duration is then decided by the propagation delay of the APD\_PULSE signal from the comparator through the FPGA logic to drive QUENCH\_ENABLE signal and AND gate output to low. In our case this is 10ns despite the physical separation of $\approx$15cm between the APD head and the SoC. The delay between quench and reset is typically set to 5ns to account for switch OFF delays of $N_1$ and $P_1$. So in figure \ref{fig:aq-anode} it can be observed that the reset is activated 15ns after GM-APD\_ANODE reaches V\_QUENCH. The reset pulse duration is kept as small as possible and is typically between 5ns-10ns depending on the value of V\_QUENCH. Table \ref{tab:reconfig-param} shows the list of configurable parameters in our design along with their range. The FPGA module for quench/reset generation runs on an internal 200MHz clock which allows to increase/decrease the timing parameters with a minimum step size of 5ns.
\begin{table}[ht]
\caption{Reconfigurable parameters and their range of supported values} 
\label{tab:reconfig-param}
\begin{center}       
\begin{tabular}{|l|c|c|c|}
\hline
\rule[-1ex]{0pt}{3.5ex}  & Min. & Max. & Units \\
\hline
\rule[-1ex]{0pt}{3.5ex} Quench duration & 10 & 1000 & ns  \\
\hline
\rule[-1ex]{0pt}{3.5ex} Reset duration & 5 & 1000 & ns  \\
\hline
\rule[-1ex]{0pt}{3.5ex} Deadtime & 35 & 1000 & ns  \\
\hline
\rule[-1ex]{0pt}{3.5ex} Quench voltage & 0 & 30 & V  \\
\hline
\rule[-1ex]{0pt}{3.5ex} Bias voltage & 0 & 500 & V  \\
\hline 
\end{tabular}
\end{center}
\end{table}
\subsection{Operation in PQAR and PQ configurations}
Our system operates in PQAR mode when the FPGA is configured to keep the QUENCH\_ENABLE signal low at all times. This disables the AND gate shown in figure \ref{aq-design} and the QUENCH signal is never triggered, keeping $N_1$ and $P_1$ always OFF. So when the APD avalanche occurs it is quenched passively through ballast resistor $R_B$ for a designated duration, after which the FPGA only drives the RESET signal to high and turns ON $N_2$. This restores the relative reverse bias voltage of the GM-APD to V\_BIAS. For passive quenching (PQ) mode, the FPGA is configured to keep both the QUENCH\_ENABLE and RESET signals low at all times. As a result $P_1,\ N_1$ and $N_2$ are never turned ON and so the APD is both quenched and reset passively. However in both modes the comparator still continues to detect APD avalanches and toggles the OUT and APD\_PULSE signals which can be counted by the FPGA.
\section{Performance characterization with commercial APD}
We characterize our system with the SAP500-T8 APD which has a built-in TEC (Thermo-electric cooler) to vary the temperature. For this APD, we present results for the lowest achievable values for deadtime, after-pulsing and operating temperature at an excess voltage of 10V for any given temperature.

%Reducing APD temperature is helpful in reducing dark counts that occur due to thermal excitation. However reducing it too much increases the after-pulsing probability. To reduce after-pulsing, the quench duration of the APD needs to be increased but this increases the deadtime and in-turn reduces the linearity and detection efficiency of the system.   
\begin{figure} [ht]
\begin{center}
\begin{tabular}{c} 
\includegraphics[width=\textwidth]{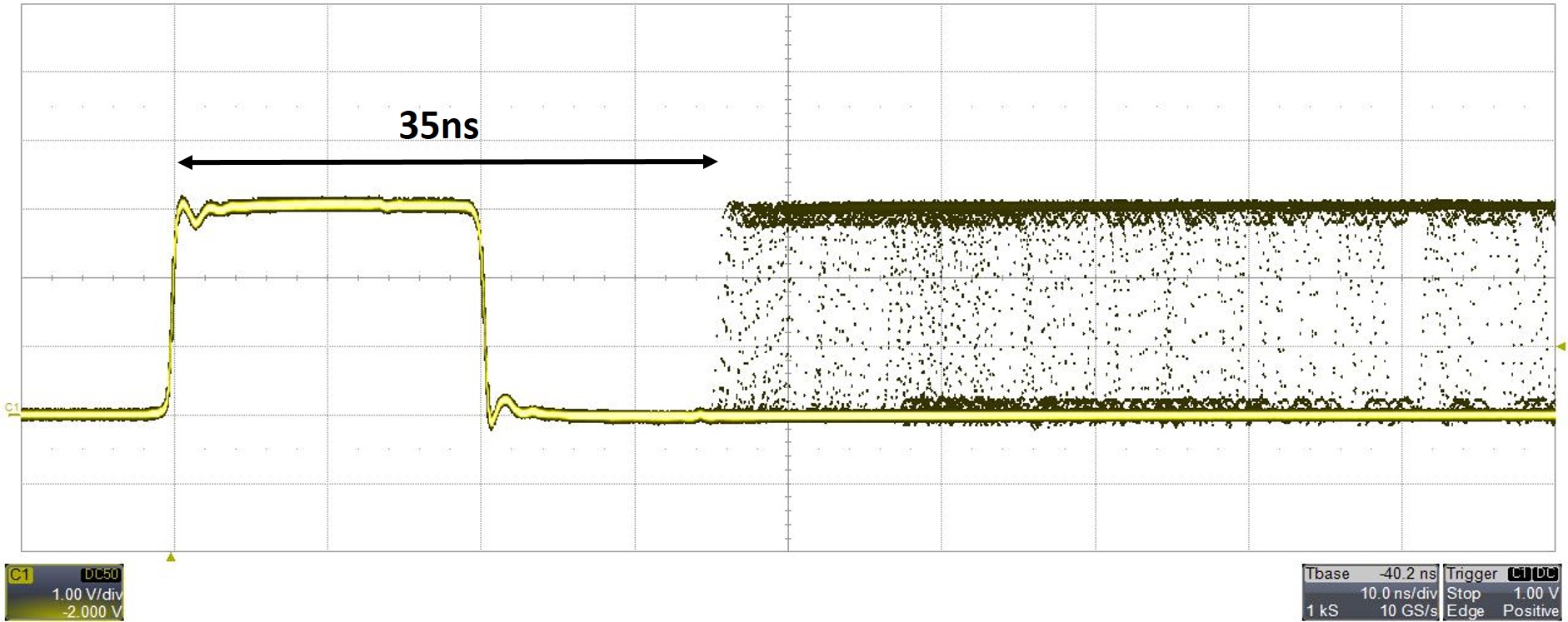}\\
\end{tabular}
\end{center}
\caption[] 
{ \label{fig:deadtime} 
Deadtime of 35ns is observed with accumulated OUT pulses (C1) captured on the oscilloscope in persistent mode at a background rate of $\approx$66Kcps}
\end{figure} 

\subsection{Deadtime}
Once an avalanche is detected, the APD cannot detect another avalanche until the first avalanche is completely quenched and the APD is reset back to its original state. This constitutes the deadtime and as shown in figure \ref{aq-timingdiagram}. 
%An important consideration in our system is the physical separation of $\approx$15cm between the APD head and the quench/reset generator logic block which is located on the SoC.To ensure this does not impact the deadtime, 
We capture the accumulated OUT pulses on the oscilloscope in persistent mode. It can be observed in figure \ref{fig:deadtime} that the deadtime (time interval between the rising edge of the first pulse and the rising edge of the earliest subsequent pulse) is 35ns which in-principle allows for detection rates of $>$28Mcps with our system. 

\subsection{After-pulsing probability and temperature}
%When the SAP500-T8 APD is operated at room temperatures 20°C to 25°C (293K-298K), the dark counts (due to thermal excitation) are higher than 20Kcps for an over-voltage of 10V. For the same over-voltage but at lower temperatures, the dark counts measured are much lesser: $\approx$5Kcps at 0°C (273K) and $\approx$4Kcps at -10°C (263K). 

%which happens when avalanche charge during a photon detection get trapped in localized defects within the APD. These charge carriers are released immediately after resetting the APD above its breakdown voltage, thus triggering another 'false' avalanche.
%This second avalanche is called an 'after-pulse' and does not occur due to an actual photon detection and such pulses are detrimental to the detection efficiency and deadtime of the system. Other than the effectiveness of the quenching circuit, the probability of after-pulsing also depends on the APD characteristics. For instance having a higher over-voltage is desirable for better detection efficiency of the APD but after-pulsing increases with the increase in over-voltage.  
 
\begin{figure} [ht]
\begin{center}
%\begin{tabular}{p{0.5\textwidth}>{c}p{0.5\textwidth}>{c}}  
%\includegraphics[height=5cm]{images/SPIE2-Min10AP-Zynq-43k-1k-17mV-35ns.pdf} & \includegraphics[height=4.65cm]{images/IPS2-AP-Zynq-43k-1k-17mV-35ns.pdf} 
\begin{tabular}{c} 
\includegraphics[width=0.5\textwidth]{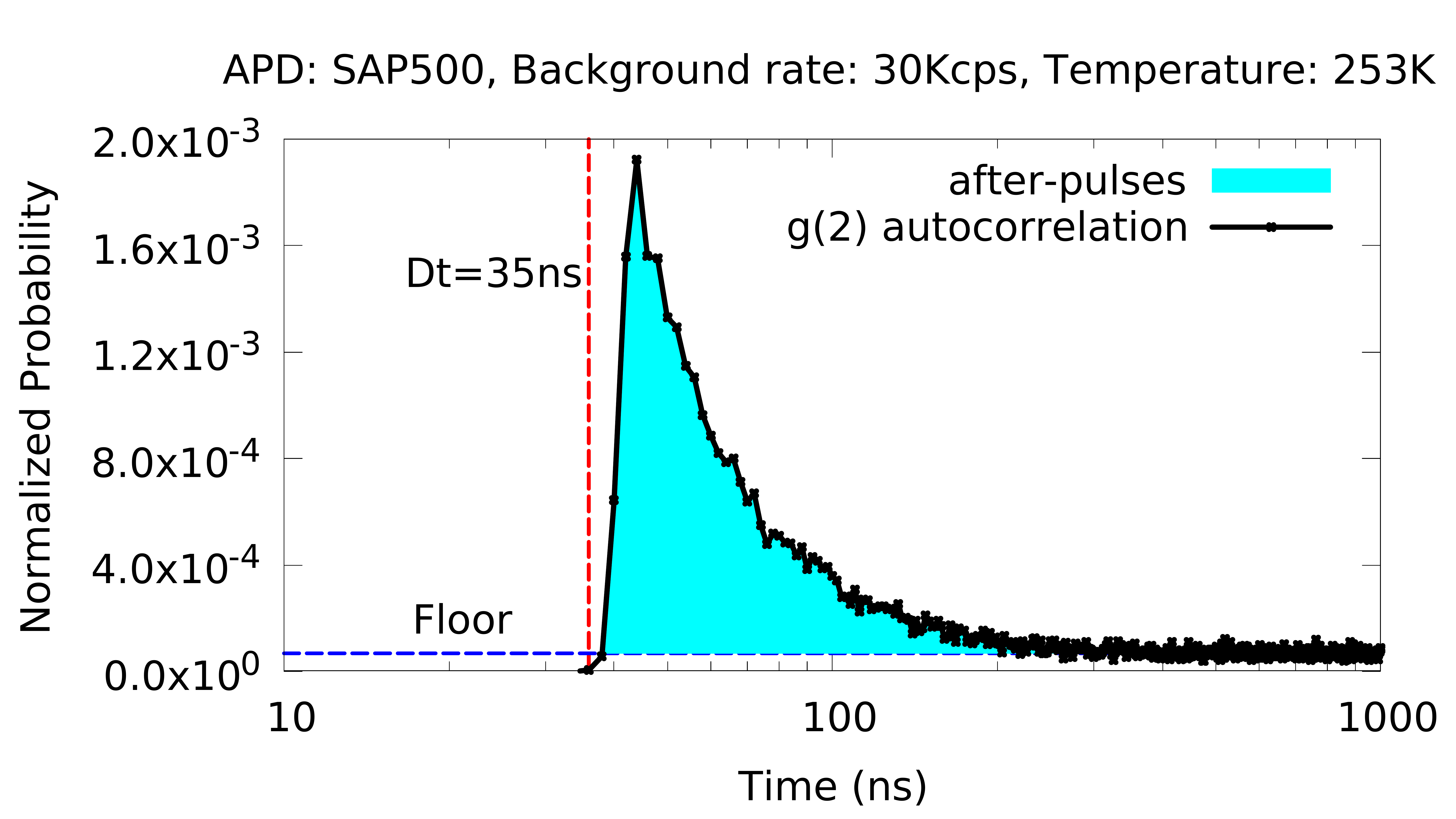}
\includegraphics[width=0.5\textwidth]{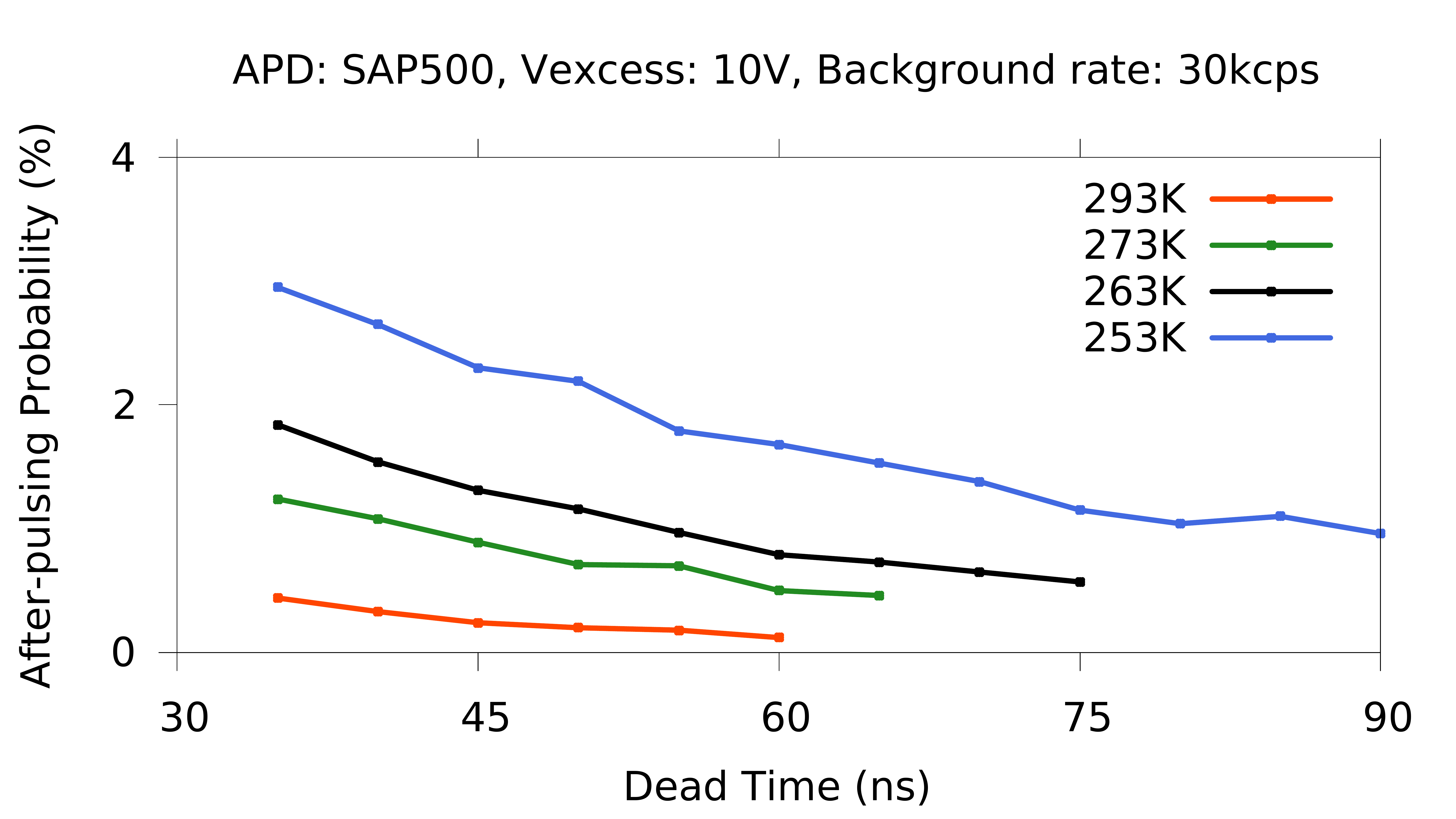}
\end{tabular}
%\begin{tabular}{p{0.5\textwidth}>{c}p{0.5\textwidth}>{c}}
\begin{tabularx}{\linewidth}{>{\centering\arraybackslash}X >{\centering\arraybackslash}X}
    (a) & (b) \\
\end{tabularx}
%\end{tabular}
\end{center}
\caption[] 
{ \label{fig:afterpulse} 
(a) shows the second order auto-correlation $g^{(2)}$ of OUT pulses when recorded with a time-tagger of 2ns time resolution. The number of after-pulses (highlighted) are higher immediately after the deadtime $Dt$ of 35ns (vertical red line) and taper off to reach a floor (horizontal blue line) after $\approx$200ns. (b) shows the variation of after-pulsing probability at different temperatures and deadtime. 
%The increase in deadtime is achieved by only increasing the quench duration while keeping other parameters constant
}
\end{figure}

It is beneficial to operate the APD at lower temperatures as it reduces dark counts (occurring due to thermal excitation) but doing so increases after-pulsing. The probability of after-pulsing also increases when the deadtime of the circuit is reduced. We measure the after-pulsing probability for the SAP500-T8 using our system for different deadtime and temperature. This is done by computing the second order auto-correlation $g^{(2)}$ of the OUT pulses using an external time-tagger module (with 2ns time resolution). One such instance is shown in figure \ref{fig:afterpulse}(a) where the deadtime ($Dt$) is 35ns, the APD temperature is 253K (-20°C) and over-voltage is 10V. The background rate is $\approx$30Kcps and the timestamps are recorded for 10 seconds which gives more than 300,000 data samples ($C_{Total}$) of arrival times in total. The $g^{(2)}$ auto-correlation is then computed at a time resolution of 2ns. In figure \ref{fig:afterpulse}(a) it can be seen that there is a high degree of correlation of arrival times immediately after the deadtime and the counts here constitute after-pulses as highlighted in the graph. After reaching a peak the counts drop exponentially until they flatten out to an average value which we call `floor', beyond which the correlation is weak or nil.
%. The flat region actually continues until 10$\mu$s, but only until 1$\mu$s is shown in figure \ref{fig:afterpulse}(a) because the correlation is weak or nil beyond $\approx$200ns and so these pulses do not count as after-pulses. 
Theoretically, the after-pulse probability $P_{ap}$\cite{Ceccarelli2019} is defined as:
\begin{equation}
    %P_{ap} = \frac{\sum_{\tau=Dt}^{+\infty}(C_{\tau}-floor)}{\sum_{\tau=1}^{\tau=T}C_{\tau}}
    P_{ap} = \frac{\sum_{\tau=Dt}^{+\infty}(C_{\tau}-floor)}{C_{Total}}
    \label{eq:afterpulse} 
\end{equation}
where $C_{Total}$ is the total number of events and $C_{\tau}$ is the count value at each time bin $\tau$. Using this approach, the percentage of after-pulsing probability for SAP500-T8 obtained with our system is shown in figure \ref{fig:afterpulse}(b) for four operating temperatures (-20°C, -10°C, 0°C and +20°C). The deadtime is varied in steps of 5ns by changing only the quench duration while keeping all other parameters constant. W.r.t. Eqn. \ref{eq:afterpulse}, for our data we consider $Dt$ as time bin of the first observed correlation event and 10$\mu$s as the upper limit for $\tau$. It is observed that $P_{ap}$ at the lowest temperature (-20°C) and lowest deadtime (35ns) is 2.95$\pm$0.08\% and remains below this value for all other settings of deadtime and temperature. For most practical applications such as QKD, it is desirable to keep the after-pulsing probability below 5\% \cite{Stipcevic17, Ceccarelli2019}.

%Linearity:
%The advantage with active quench circuits is their high detection rate and it is crucial that the circuit detects most of the incident photons. This is termed as linearity and is typically the ratio of number of detected pulses versus number of incident pulses. In reality, no circuit will be able to detect 100\% of all incident photons, however the higher the number the better is the linearity. The linearity of our implementation was tested with an optical set up that generates correlated photon pairs using Spontaneous parametric down-conversion [3] and the linearity was found to be around 87\% for a detected rate of 1M counts per second. This means that our system could detect 1 million of the 1.15 million incident photons.

\section{Integration with custom chip-scale APD}

\begin{figure} [ht]
\begin{center}
\begin{tabular}{c} 
\includegraphics[height=8cm]{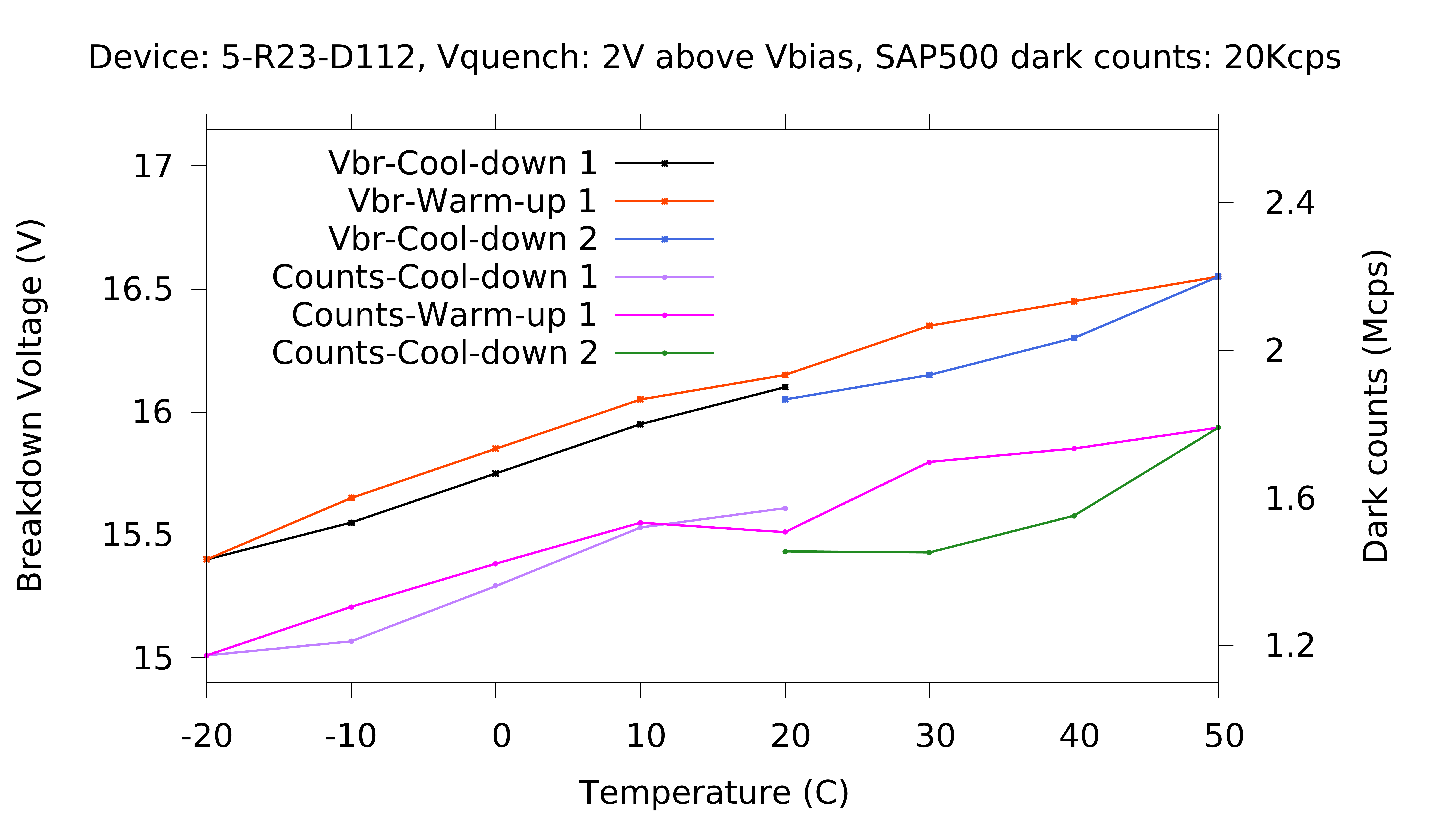}
\end{tabular}
\end{center}
\caption[] 
{ \label{fig:astar-data} 
Variation of breakdown voltage and dark counts w.r.t. temperature for one of the chip-scale APDs in Geiger mode, obtained using our SoC based active quench system. Both data plots were obtained in three consecutive phases in which the APD was first cooled down to -20°C, then warmed up to +50°C and again cooled back to +20°C}
\end{figure} 
After validating the performance of our active quench system with a widely used commercial APD (SAP500-T8), we further use it for testing our custom fabricated chip-scale APDs\cite{imreAPD21} in Geiger mode. These are Si waveguide APDs integrated with on-chip photonic waveguides and are interfaced with the active quench system using RF probes. The active quench system shown in figure \ref{aq-design} was modified to include a second identical active quench channel so that both the SAP500 and custom chip-scale APD can be simultaneously connected and operated in parallel. Both GM-APDs are controlled by only one SoC through replication of the quench/reset generator and the counter modules on the FPGA. The SAP500 was operated at room temperature while the temperature of the chip-scale APD was varied between -20°C and +50°C using a TEC. To rule out the presence of any stray light in the setup, the count rate of the SAP500 was monitored closely during the experiments to ensure that the dark counts always remained at $\approx$20Kcps, which is the expected value for this APD at room temperature when operated at an excess voltage of 10V. The breakdown voltage of the chip-scale APDs are typically between 16V to 25V and their V\_BIAS is derived from a bench-top power supply. To  prevent the breakdown voltage of the chip-scale APDs from drifting \cite{imreAPD21}, V\_QUENCH was maintained between 18V to 27V in conjunction with V\_BIAS, so as to always keep it 2V above V\_BIAS. Although V\_QUENCH for both APDs was derived from the same bench-top power supply, any changes in V\_QUENCH does not affect the operation of the SAP500 since V\_QUENCH is always greater than 15V for all tests. Considering these high voltage values, the durations of quench, reset and deadtime for both channels were set conservatively higher so as to provide sufficient time for the MOSFETs to switch ON/OFF.
%Having a bigger quench duration also ensures the after-pulsing of the APDs stays quite low even at lower temperatures. 
So for both channels, the quench duration was set to 25ns, the reset duration to 15ns and a delay of 10ns between quench and reset, giving a total deadtime of $\approx$65ns (inclusive of initial response time and all other delays) for all experiments. V\_QUENCH and V\_BIAS of the chip-scale APD were changed at run-time using scripts running on a laptop PC, while the other parameters in Table \ref{tab:reconfig-param} remained fixed.

%which can be configured at run-time by commands from the PC over a USB interface. This enabled to automate the process of varying both voltages using scripts running on the PC. The associated driver software for the TEC module was used for changing the temperature, which was also configurable through scripts on the PC. Using the described setup we measured breakdown voltage and dark counts of the chip-scale APD for varying temperatures. 
Figure \ref{fig:astar-data} shows variation in dark count and breakdown voltage w.r.t. temperature for one chip-scale APD. To check for hysteresis in breakdown voltage, the temperature of the APD was changed in three continuous phases. It was first cooled from +20°C to -20°C, then warmed up from -20°C to +50°C and again cooled down from +50°C to +20°C. In each phase temperature was changed in steps of 10°C. At each temperature, V\_BIAS was initially set to a value well below the breakdown voltage and the detected counts were ensured to be 0. V\_BIAS was then increased in steps of 0.05V every 2 seconds until the SoC started measuring counts, which indicated the breakdown voltage. V\_BIAS was then kept constant and the dark counts were recorded. Before each new measurement the APD was reset by switching off V\_BIAS. 
%At the initial temperature of +20°C, the breakdown voltage for this device was measured to be 16.1V and the dark count obtained was $\approx$1.57Mcps. 
It can be observed that the variation of breakdown voltage and dark count w.r.t. temperature follows a well known trend as seen with existing commercial devices.

\subsection{Advantage of using SoC based active quenching for testing chip-scale APDs}
For preliminary tests, only dark counts and breakdown voltage of one chip-scale APD have been recorded. Future tests will involve characterizing the after-pulsing probability (like shown in figure \ref{fig:afterpulse}) of each APD for at least 10 different deadtimes, 5 different temperatures and 3 different excess voltages, which amounts to a minimum of 150 test cases per APD. There are more than 400 variants of our custom chip-scale APDs with each differing in channel length, width, doping concentration etc. So characterizing after-pulse probability for all of them requires 60,000 tests in total. This means the active quench system needs to be re-configured 60,000 times for executing the complete test set. Using hard-wired fixed active quench circuits for this purpose is impractical since it requires a lot of manual intervention for changing parameters, which is both laborious and error prone. Using our SoC based active quench system instead paves way for making these adjustments instantly with script-based automation thus saving both cost and time.

%are fabricated on a single chip containing several variants in which each APD differs from the other in terms of channel length, width, doping concentration, etc. As shown in figure \ref{fig:astar-apdarch}, there exist two physical architectures of our custom chip-scale APDs -  those with wire-bonded dual-in-line packaging (DIP) and those which are bare without any packaging and interconnect, .  
%This is especially true for future testing with our custom chip-scale APDs, for which there exist several hundred variants with each one different from the other in terms of channel length, width, doping concentration, etc.  This results in an exponential increase in the total number of test cases and  
\section{Conclusion and future work}
We have described a hybrid design for a re-configurable active quench system which can be used to characterize multiple APD architectures. We have implemented the design with readily available components and characterized it with the SAP500 commercial APD cooled to -20°C and achieved a deadtime of 35ns while limiting the after-pulsing probability to $\approx$3\%. We have also integrated our active quench system directly with custom chip-scale APDs and presented preliminary results for variations in their breakdown voltage and dark counts w.r.t. temperature. Our active quench system can also be used in other experiments and commercial applications such as Quantum information processing, random number generators, time-correlated single photon counting (TCSPC) and range measurement applications such as LIDAR. Looking ahead, the deadtime of the quenching system can be reduced further by using faster MOSFETs which will increase the detection rate and efficiency. The SoC firmware can be upgraded to directly compute after-pulse probability without additional equipment. 

\section{Acknowledgements}
%This research is supported by National Research Foundation, Singapore under the grant NRF-CRP12-2013-02 [?] - inputs needed
%This work is supported by National Research Foundation, Singapore under its Central Gap Fund (NRF2018NRFCG001\-001)
This research is supported by the National Research Foundation, Singapore under its Central Gap Fund (NRF2018-NRFCG001-001)

% References
\bibliography{Bib-ReconfAQ} % bibliography data in report.bib
\bibliographystyle{spiebib} % makes bibtex use spiebib.bst

\end{document}